\documentclass[11pt]{article}
\usepackage{amsmath}
\usepackage{graphicx}
\usepackage{latexsym}
\usepackage{amssymb,amsthm}

\theoremstyle{definition}

\numberwithin{equation}{section}

\begin{document}

\title{Classical quarks in dual electromagnetic fields}
\author{Harry Schiff\\Professor Emeritus at
              Department of Physics \\
University of Alberta\\
Edmonton, AB, Canada, T6G 2J1\thanks{Mailing address:  304--2323 Hamiota St. Victoria, BC. CA V8R 2N1;   email: hschiff@shaw.ca.}}

\date{}

\maketitle

\begin{abstract}
Electromagnetic properties of quark-like particles are examined in a classical field model involving extended dual electromagnetic fields. These can have fractional charges and a confining potential that derives essentially completely from a short-range weaker potential. The combined potentials exhibit an asymptotically free spherical surface and contribute to the masses of the particles. The quarks are shown to have an intrinsic symmetry that describes their structures in hadrons. Multi-quark solutions are easily obtained for both stable and unstable particles. Each quark can undergo simple harmonic motion in a range of  frequencies.

\end{abstract}

 \noindent PACS numbers:  12.39.-x

\section{Introduction}

In this note a classical electromagnetic model of particles is introduced that have a number of quark-like properties. Starting with the Li\'{e}nard-Wiechert exact solutions of Maxwell's equations\footnote{See, for example, J. D. Jackson, Classical Electrodynamics, (John Wiley \& Sons, Inc. 1975, 2nd ed) p.\ 654 ff.} of a point charge $q$ in arbitrary motion, it can be shown that the following relation between the fields and potentials holds everywhere:  ${\bf [e = c= 1]}$
\begin{equation}		
(B^2 - E^2)  =  -q^{-2}(A^2 - \phi ^2)^2   		
\end{equation}
With $F_{\mu \nu}  = \partial_\mu A_\nu - \partial _\nu A_\mu , 3$-potential vector ${\bf A}, A_4 = i\phi$ and the  relation 
$F_{\mu \nu} F_{\mu \nu}  = 2(B^2 - E^2)$, (1.1) can be written
\begin{equation}
F_{\mu \nu}  F_{\mu \nu}  = - 2 q^{-2} (A _\mu A _\mu ) ^2 
\end{equation}	
The Li\'{e}nard-Wiechert solutions also satisfy the orthogonal relation $\mathbf{E} \cdot \mathbf{B} = 0 $ everywhere, i.e.,
\begin{equation}
{}^\ast\!F_{\mu \nu} F_{\mu \nu}  = 0 
\end{equation}	
 In (1.3) the familiar notation  ${}^\ast\!F_{\mu \nu}    =   1/2 \,  i \varepsilon _{ \mu \nu \alpha \beta }  F_{\alpha \beta} $ is used for the dual of $ F_{\mu \nu} $  using the Levi-Civita tensor $ \varepsilon _{ \mu \nu \alpha \beta }$.		
Due to the opposite symmetries of $ F_{\mu \nu} $  and $ A _\mu A _\nu $, (1.2) can be written,	
\begin{equation} ( F_{\mu \nu} + \surd {2} q ^{-1} A_\mu A _\nu )^2 = 0 
\end{equation}
Defining the mixed tensor 
\begin{equation}
W_{\mu \nu} \equiv F_{\mu \nu} + \surd 2 q ^{-1} A _\mu A _\nu , 
\end{equation}
(1.4) can also be written 
\begin{equation}
W_{\mu \nu} W_{\mu \nu} = 0 
\end{equation}
Further, noting that $ {}^\ast\!W_{\mu \nu} = {}^\ast\!F_{\mu \nu} $,  since  $ \varepsilon _{\mu \nu \alpha \beta }A _\alpha A_\beta   =  0$, (1.3) can be written,
\begin{equation}{}^\ast\!W_{\mu \nu} W_{\mu \nu} = 2 \mathbf{E} \cdot \mathbf{B} = 0 \end{equation}
The expressions (1.6) and (1.7) show that $W_{\mu \nu}$ (1.5) satisfies formally the conditions for a null field (although the term `null field' applies strictly to an anti-symmetric tensor we have taken the liberty to use it here for the mixed tensor)	 that holds everywhere in space, as compared for example to solutions $F_{\mu \nu}$  to Maxwell's equations where, in general, only the radiation field is a null field. This interesting property of $W_{\mu \nu}$ a combination of the exact solutions for the electric and magnetic fields and potentials of Maxwell's equation for a point charge, arouses curiosity as to whether the expression $W_{\mu \nu}$ may exhibit other features of interest. To this end we consider the non-gauge invariant equation $\partial _\mu W_{\mu \nu} = - 4 \pi  {}^e\!\!j _\nu$,  
\begin{equation}
\partial _\mu (F_{\mu \nu} + \surd 2 q ^{-1} A_\mu A_\nu ) = - 4 \pi {}^e\!\!j _\nu 
\end{equation}				 
where ${}^e\!j_\nu$ is an external localized 4-current density $({}^e\!{\bf J}, i{}^e\!\!\rho)$ that can be chosen freely.  
The total current density thus consists of the localized external current ${}^e\!j_\nu$ density plus the self-field contribution $\surd 2  (  4 \pi q ) ^{-1} \partial _\mu (A _\mu A_\nu)$. 
 \section{Solutions of (1.8)} 

We will examine static, radially symmetric solutions of (1.8),
\begin{equation}
\surd 2 q ^{-1} \nabla \cdot (A{\bf A}) = - 4 \pi {}^e\!\!J(r)
\end{equation}				 				
\begin{equation}
 \nabla \cdot {\bf E} - \surd 2 q^{-1} \nabla \cdot (\phi {\bf A}) = 4 \pi {}^e\!\!\rho (r) 				 
\end{equation}
The asymptotic solution for (2.1),
\begin{equation}
\nabla \cdot(A{\bf A}) = 0 ,	
\end{equation}		
\begin{equation}
A = a/  r 	\;\;\;
  \mbox{(a  arbitrary)}
\end{equation}
Defining	
\begin{equation}    
\gamma \equiv \surd 2 q^{-1} a , 	
\end{equation}			
\begin{equation}
A = \gamma q / \surd 2 r 
\end{equation}	
Asymptotically, using (2.5), (2.2) can be written,
\begin{equation}
\frac{d^2 (r \phi) }{dr^2} + \frac{\gamma d ( r \phi) }{rdr} = 0 
\end{equation}
For any value of $ \gamma$, except $\gamma = 1$, treated separately below, there are two indicial solutions of (2.7),
\begin{equation}
\phi _1 = b /r
\end{equation}
\begin{equation}				 
\phi_2 = c r ^{-\gamma}
\end{equation} 	
A Coulomb potential and a possible absolute confining potential for $\gamma < 0$, where $c$ may be positive or negative. For any value of $\gamma$ the total charge of the external source in (2.2), using the divergence theorem, 
\begin{equation} \int {}^e\!\!\rho d ^3 x = (1 - \gamma ) b 
\end{equation}

	For $\gamma =1$ the two indicial solutions of (2.7) merge to a single Coulomb potential so a second solution is needed. This is easily seen to be given by $r\phi  \sim  \log r$. We note that for $ \gamma =1$ the total external charge is zero and one can choose its charge density to be zero. These solutions however will not be used here. In the following the particular solution to (2.2) for which (2.8) is the asymptotic solution will be called $\phi_p$.

Using the divergence theorem in (2.1) the total 3-current of the source is
\begin{equation}
\int {}^e\!J (r) d^3 x = - 1/4 \pi \int \surd 2 q ^{-1} A ^2 r^2 d \Omega = - \gamma^2 q / \surd 2 
\end{equation}

\section{Null Condition and Fractional Charges}

	As $A$ and $\phi$   go as $1/r$ asymptotically, the application of (1.2) is suggested to
these solutions,  leading to an equation for the charge $b$.
 Thus, with $B = 0$, (1.2) becomes
\begin{equation}
- E ^2 + q ^{-2} (A ^2 - \phi ^2) ^2 = 0 					
\end{equation}
For each sign of $q$ one obtains a quadratic equation for the Coulomb charge $b$ in (2.8),
\begin{equation}
	 b^2 \pm q b - \gamma ^2 q^2 / 2 = 0 				
\end{equation}
Solutions $b_-$, with the minus sign in (3.2), have opposite signs to those with the plus sign $b_+$, equivalent to $q \rightarrow -q$. Thus for the quadratic with the minus sign,
\begin{equation}
b^2 - q b - \gamma ^2 q ^2 /2 = 0 ,					
\end{equation}
$b_-$  has the two solutions, 
\begin{equation}		
b_- (\pm) = q [1 \pm \surd (1 + 2 \gamma ^2 )]/2 			
\end{equation}
In (3.4) choosing $\gamma^2 = 4$ and $q = \pm 1/3 $, one gets 
\begin{equation}				
b_- (\pm)  = \pm (2/3, - 1/3 ), \; b _+ = - b _-				
\end{equation}
 (With $b = \pm 1/3, \pm 2/3$ and $\gamma = -2$ (negative for confinement), the external charge (2.10) is $\pm 1, \pm 2$ respectively. Other possible fractional values for $b$ can be obtained, for example with $\gamma^2 = 4$ and 
$q = \pm 2/3, b_- = \pm (4/3, -2/3) , b_+  = - b_-)$.

One may also note that with $\gamma^2 = 4$, choosing only the minus sign on the right hand side of (3.4) with $q = \pm 1$ one gets the two values $b = -1$ and $+1$. 

\section{Lagrangian}

A Lagrangian for (1.8) cannot be obtained using only the field components in that equation. 
The nonlinear addition to the second order Maxwell terms of the potentials is in first order, a problem similar to that of Newton's equations of motion with a first order dissipative term. This is resolved for (1.8) with the addition of an electromagnetic field\footnote{See, for example, Morse \& Feshbach, Methods of Theoretical Physics (McGraw-Hill Book Company, Inc.,  1953), Vol. 1, p.\ 298, 313.  Where they discuss introducing mirror image systems for dissipative systems.}  $G_{\mu \nu}$   and its associated potential $V_\mu$,
\begin{equation}
 G_{\mu \nu } = \partial _\mu V _\nu - \partial _\nu V _\mu 		
\end{equation}
The following Lagrangian includes  $ F_{\mu \nu} , A _\nu $ and $ G_{\mu \nu}, V_\nu $ with their respective currents ${}^e\!j_\nu$ and ${}^\ast\!\!j_\nu$. 
\begin{equation}
   L = - 1/8 \pi [ 1/2 G_{\mu \nu} F_{\mu \nu} + \surd 2 q ^{-1} V_\nu \partial _\mu (A _\mu A_\nu ) + 4 \pi V_\nu {}^e\!j _\nu + 4 \pi A_\nu {}^\ast\!\!j _\nu ]		 
\end{equation}
Variation of $L$ with respect to $V_\mu$ gives (1.8), while variation with respect to $A_\mu$ yields a linear equation for $V_\mu$. 
\begin{equation} 
   	\partial _\mu G_{\mu \nu} - \surd 2 q ^{-1} A_\mu (\partial _\mu V_\nu + \partial _\nu V_\mu ) = - 4 \pi^\ast\!\!j_\nu
\end{equation}
The external source may be chosen freely.  The total current density
\begin{equation}
j_\nu = - \surd 2 (4 \pi q) ^{-1} A_\mu (\partial _\mu V_\nu + \partial _\nu V_\mu ) + {}^\ast\!\!j _\nu 
\end{equation}
is conserved, and for radial solutions considered here the total 3-current density $J = 0$.

From (4.3) the equation for $V_4 = i \psi$, becomes  
\begin{equation}			
\frac{d^2 \psi}{dr^2} + \frac{2d \psi}{rdr} - \surd 2 q^{-1} A \frac{d \psi}{dr} = - 4 \pi {}^\ast\!\!\rho (r) 
\end{equation}		
Both particular and homogeneous solutions of (4.5) will be considered. Before dealing with general solutions to (2.2) and (4.5) in Section 6, it is useful to look first at an explicit solution for a single-quark. First, for the homogeneous function $\psi_h$ of (4.5), replacing  $ \surd 2 q^{-1} A  $ by $  A^\ast $,  
\begin{equation}
 \frac{d^2 \psi_h}{dr^2} + \frac{2d \psi_h}{rdr} -   A^\ast  \frac{d \psi_h}{dr} = 0 				
\end{equation}
Using the asymptotic value of $A$ in (2.6) with $\gamma  = -2$, (4.6) becomes,
\begin{equation}		  			  	                                         
 \frac{d^2\psi_h}{dr^2} + \frac{4d\psi _h}{rdr} = 0                                   	
\end{equation}
It follows that  for all asymptotic $ \psi_h$  solutions,
\begin{equation}
\psi_h \sim 1 / r ^3 +   \mbox{const.}	         
\end{equation}
To find $A^\ast$ in (4.6) we choose for $\psi_h$, to be defined as $\psi_\lambda$, the simple monotonic function 
\begin{equation}
\psi_h \equiv \psi_\lambda = \lambda s^2 /(s  + \alpha r)^3 + \lambda s^2 t
\end{equation}
 $\lambda$ is a charge plus an amplitude, $s$ is a scale parameter and $\alpha$ a multiplier for separating distance between $s $ and $r$. An arbitrary constant $t$ has been added to $\psi_\lambda$  with dimension $1/s^3$. (One expects $s$ to be small compared to the position of the quark, i.e., $ \ \alpha \ll 1$). Using (4.9) one finds for the 3-potential $A^\ast$ in (4.6), 
\begin{equation}	
 A^\ast = 2 / r - 4 \alpha /(s + \alpha \, r)					 
\end{equation}
 With (4.10) in (2.2), which is a complete divergence,  putting the divergence constant $K$  in $K/r^2$  equal to zero, one obtains the following homogeneous solution for the confining potential, 
\begin{equation}
  \phi _k = k (s + \alpha r )^4 /s^3 r^2 	
\end{equation}
$k$ is a charge plus an amplitude. $\phi_k$ corresponds asymptotically to (2.9) ($ k/s^3$ replaces $c$) 
and diverges as $1/r^2$ at the origin with a minimum value of $16k\alpha^2/s$ at $\alpha r  = s$. 

Substituting (4.10) in (2.1) one finds for the external 3-current density
\begin{equation}
  {}^e\!\!J(r) = - 2 \surd 2 q [\delta (r) - \alpha ^2 s (s -  \alpha r) /\pi r^2 (s + \alpha r)^3  ] 			 
\end{equation}
Spatial integration of (4.12) is in accordance with (2.11).

	For the 3-potential $V$ in (4.3) one can choose the external 3-current density ${}^\ast\!\!J = 0$, the equation for the $V$ would then become,  
\begin{equation}   
2 A^\ast \frac{dV}{dr} = \frac{\phi d \psi}{dr}                                                            
\end{equation}	
However, since $V$ will have no bearing on the physical quantities we consider, its solution will 
not be of interest. For the particular solution $\phi_p$ of (2.2) we are free to choose, 
\begin{equation}
\phi_p = b \alpha ^3 r^2 / (s + \alpha r)^3 
\end{equation}
For energy considerations later the particular solution $\psi_p$ of (4.5) is taken to be the same function as (4.14),
\begin{equation}
\psi _p = \phi_p = b \alpha ^3 r^2 /(s + \alpha \, r)^3					
\end{equation}
In (4.5), as $\psi_p, \phi_p$ and $A^\ast$ go as $1/r$ asymptotically, both the field and external charge densities are seen to have a $1/r^3$ singularity with opposite signs that cancel for a non-singular total charge density.

\section{Energy and Mass}

From the Lagrangian (4.2) the canonical stress-energy tensor is
\begin{equation}
T_{\alpha \beta} = - \frac{\partial L}{\partial (\partial _\alpha A _\lambda )} \partial _\beta A_\lambda -\frac{\partial L}{\partial (\partial _\alpha V _\lambda )} \partial _\beta V_\lambda + L \delta _{\alpha \beta} 
\end{equation}

To represent the energy of a particle, we choose for simplicity, the stress-energy tensor of the electromagnetic fields of (5.1),  $F_{\alpha \mu  }$ and $G_{\alpha \mu }$,
\begin{equation}
\Theta_{\alpha \beta} = 1/8 \pi [ G_{\alpha \mu } F_{\beta \mu } + F_{\alpha \mu } G_{\beta \mu } - 1/2 (G_{\mu \nu} G_{ \mu\nu } ) \delta _{\alpha \beta } ]
\end{equation}
Further, in (5.2) only the particular solutions $\psi_p$ and $\phi_p$ are chosen for $G_{\alpha \mu }$ and $F_{\alpha \mu}$ and as these solutions are now equal, $G_{\alpha \mu} = F_{\alpha \mu}$. This leads to a familiar expression for this energy density,
\begin{equation}
-\Theta _{44} = 1/8 \pi [ -2F_{4 \mu } F_{4 \mu } +   1/2 (F_{\mu \nu} F_{ \mu\nu } ]= 1/8 \pi ( d \phi_p/dr)^2 
\end{equation}	
with integrated Coulomb energy
\begin{equation}
\xi _c = 1/2 \int( d \phi _p / d r) ^2 r ^2 dr 
\end{equation}

For the contributions to the energy by the homogeneous potentials $\phi_k$ and $\psi_\lambda $ we choose these energies to be the interactions of a quark charge $b$ with the potentials, $b\phi_k$ and $b\psi_\lambda $.  Thus, when added to (5.4), the total energy for the single-quark, (4.9) and (4.11), becomes,
\begin{equation}
\xi = \xi_c + \xi_k + \xi _\lambda = 0.043 \alpha b ^2 /s + b k (s + \alpha r )^4 /s^3 r^2 + b \lambda s^2 /(s + \alpha r)^3 + b \lambda s ^2 t 
\end{equation}
For the numerical value in (5.5) the function $\phi_p$ in (4.14) was used. For the mass one needs the minimum of $ \xi_k + \xi _\lambda $ plus the Coulomb energy. In general, the result for the minimum will depend on $k, \lambda $, and $\alpha $ and will be different from the $\xi_k$ minimum---such as $16bk\alpha^2/s$ at $\alpha r  =  s$ for the single quark. The spherical surface corresponding to the minimum energy is evidently one of `asymptotic freedom'. The energies and masses for the single-quark and multi-quarks are dealt with in the next section.

\section{Multi-Quark Solutions}

General solutions for $\phi_k$ in (2.2) requires $A^\ast$ from (4.6),
\begin{equation}
A^\ast = \psi _\lambda ^{\prime \prime} / \psi _\lambda ^\prime + 2/r,
\end{equation}			
The dimensional constants $\lambda s^2$ cancel in $\psi _\lambda ^{\prime \prime }/\psi _\lambda ^\prime$, replacing $\psi _\lambda/\lambda s^2$ by the capital $\Psi_\lambda$, one can write
\begin{equation}
\psi _\lambda ^{\prime \prime}/ \psi _\lambda ^{\prime }= \Psi _\lambda ^{\prime \prime} / \Psi _\lambda ^{\prime }  = d \log (| \Psi _\lambda ^{\prime } |) /dr 
\end{equation}
and	
\begin{equation}
A^\ast = d \log (r^2 | \Psi _\lambda ^{\prime } |) /dr 			
\end{equation}
In (2.2), with the equation for $\phi_k$  a complete divergence, replacing $\phi_k/(k/s^3)$ by $\Phi_k$,
\begin{equation}
d \Phi_k / dr + \Phi _k A ^\ast = {K} s^3 /r^2 
\end{equation}
In (6.4), $s^3$ has been added for a dimensionless ${K}$. With $A^\ast$ from (6.3), (6.4) 
becomes					 	
\begin{equation}
d \Phi_k / dr + \Phi _k [d (r^2 | \Psi _\lambda ^\prime |) / dr ] / r^2 | \Psi _\lambda ^\prime | = {K} s^3 /r^2 
\end{equation}
Multiplying by $ r^2  | \Psi _\lambda ^\prime | $,
\begin{equation}
r^2  | \Psi _\lambda ^\prime | d \Phi_k / dr + \Phi_k d (r^2  | \Psi _\lambda ^\prime | )/dr = {K} s^3    | \Psi _\lambda ^\prime | 
\end{equation}
\begin{equation}
d (\Phi_k r^2 |\Psi _\lambda ^\prime | )/dr = {K} s^3    | \Psi _\lambda ^\prime | 
\end{equation}
\begin{equation}
\Phi_k = {K} s^3 \left[\int | \Psi_\lambda ^\prime | dr \right] / r^2 | \Psi _\lambda ^\prime | + C /r^2 |\Psi _\lambda ^\prime |
\end{equation}
With 				
\begin{equation}
\left[ \int |\Psi _\lambda ^\prime | dr\right] /r^2 |\Psi _\lambda ^\prime | = \left[ \int \Psi _\lambda ^\prime dr\right] /r^2 \Psi _\lambda ^\prime 		 			
\end{equation}
\begin{equation}
\Phi_k = {K} s^3 ( \Psi_\lambda + \beta ) /r^2 \Psi _\lambda ^\prime  + C /r^2 |\Psi _\lambda ^\prime |
\end{equation}

The constant $\beta$ (dimension $1/s^3$), consists of the integration constant plus the arbitrary constant added to $\Psi_\lambda $. Putting ${K}s^3\beta  = P$, one obtains the three distinct terms,	
\begin{equation}	
\Phi_k = {K} s^3 \Psi _\lambda  /r^2 \Psi _\lambda ^\prime + P /r^2 \Psi _\lambda ^\prime + C /r^2 |\Psi _\lambda ^\prime|	
\end{equation}			
We note that the $K$ and $P$ terms change sign in crossing an extremum of $\Psi _\lambda $. With different $\Psi _\lambda $ various multi-quark solutions follow from this general solution to the confining potential, defined by the relative signs and magnitudes of the three dimensionless parameters $K, P$ and $C$. Some examples are shown below.

From (6.3), asymptotically $A ^\ast \rightarrow -2/r$ and near $r = 0, A^\ast \rightarrow  (n +1)/r$ for $\Psi _\lambda \sim  r^n, n\geq 1$, between these 
points $A^\ast$ diverges at the extrema of $\Psi _\lambda $. Near the origin in  (6.4), with The K term from (6.11), one finds readily the necessary equality  with    $K/r^2$;  including all term in (6.11), $\Phi_k$ goes as $\pm 1/r^{n + 1}$ and diverges at the extrema of $\Psi _\lambda $. Asymptotically the ${K}$ term goes as $\pm 1/r$ and the $P$ and $C$ terms go as $\pm  r^2$. We will consider a few choices for the magnitudes and signs of the three parameters $K, P$ and $C$, for a variety of hadrons.  To summarize, for all solutions of (6.11) and (6.3):
\begin{eqnarray}
&&\mbox{\rm (a)}\quad 	 \Phi_k \sim \pm r^2 \;\;\mbox{asymptotically}\nonumber\\
&&\mbox{\rm (b)}	\quad \Phi_k \sim \pm 1/r^{n+1} \;\;\mbox{near the origin},\;  n \geq 1	\nonumber\\	 
&&\mbox{\rm (c)}\quad 	\Phi_k   \; \mbox{and}\; A^\ast   \;\;\mbox{diverge at the extrema of}\;  \Psi _\lambda  \\
&&\mbox{\rm (d)}	\quad A^\ast  \rightarrow -2/r  \;\;\mbox{asymptotically and as}\;  (n +1)/r \;\;\mbox{near the origin}\nonumber
 \end{eqnarray}
The behavior in (6.12) is readily seen to apply to the single-quark solutions (4.9), (4.10) and (4.11), except for (c), as (4.9) has no maximum or minimum. 

For the simplest case in (6.11), where ${K} = 0$, 
\begin{equation}
\Phi _k = C / r^2 | \Psi _\lambda ^\prime |						
\end{equation}
From (6.3) and (6.13),							
\begin{equation}
A^\ast = - (1 / \Phi _k ) d \Phi _k / dr 
\end{equation}
In this case $A^\ast = 0$ at the extrema of $ \Phi_k$.

	For a 2-quark system a function $ \psi_\lambda $  is needed with one maximum. The $\phi_k$ divergence at this maximum (6.12c) divides space into two regions in each of which $\phi_k$ is contained by two divergences, between the origin and the central divergence in the first region and between the central divergence and asymptotically in the second region. As an example of a simple $ \psi_\lambda $  with one maximum, applied to a charged 2-quark system, we consider
\begin{equation}
\psi_\lambda = \lambda s^2 r /(s + \alpha r)^4  			
\end{equation}
\begin{equation}	
\psi_\lambda ^\prime = \lambda s^2 (s- 3 \alpha r)/( s + \alpha r)^5 		 			
\end{equation}
\begin{equation}	
\phi_k = k /s^3 (s + \alpha r )^5 /r^2 |s - 3 \alpha r|				
\end{equation}

In (6.17), $\phi_k$ was obtained using (6.13), with $C$ included with the amplitude in $k$. At the maximum of $d\psi_\lambda/dr$, where $r  = s/3\alpha$, the $\phi_k$ divergence there divides the space into the two regions, $
r \leq s/3\alpha $  and $r \geq  s/3\alpha $. In region(1), $\phi_k$ has a divergence at $r = 0$ and at 
$r = s/3\alpha $, while in region(2), $\phi_k$ has a divergence at $r = s/3\alpha $,
and asymptotically. The zero derivative of (6.17) results in a quadratic for the two $\phi_k$ minima, $r_{km} = 0.18s/\alpha$ and $1.82s/\alpha$, straddling the divergence at $r =s/3\alpha$. In each region the quark mass, with appropriate $b$, obtains 
from the minimum of the energy $b(\phi_k + \psi_\lambda )$.

In the region of a quark in a multi-quark system, $d\psi_k/dr$ is either positive or negative, so the energy minimum there, from $d[b(\phi_k+ \psi_\lambda )]/dr = 0$, can be written,   
\begin{equation}
d\phi_k /dr = - d 	\psi_\lambda /dr = \pm |\psi_\lambda^\prime |,	
\end{equation}		 				 
where the $+$ or $-$ signs refer to $\psi_\lambda^\prime < 0$ or $> 0$ respectively. With $ K$ small enough to be ignored, introducing the dimensional factors and using (6.13), (6.18) can be written,
\begin{equation}
d \Phi_k /dr = \pm s^5 \lambda | \Psi_\lambda^\prime |/k = \pm s^5 \lambda /k r ^2 \Phi_k
\end{equation}
\begin{equation}
\Phi_k d \Phi_k /dr = \pm s^5 \lambda /k r^2 
\end{equation}			        		 
	
(To illustrate, we take $k > 0, \lambda > 0)$. An $r$-value, $r_m$, that satisfies (6.20) minimizes the energy $\xi_k+ \xi_\lambda $  between two divergences. When $\Psi_\lambda^\prime < 0$ and $d\Phi_k/dr  > 0$, the $1/ r^2$ curve in (6.20) intersects the right half of $\Phi_k$ and the actual minimum, $r_m$, for that quark moves to the right of the $\Phi_k$ minimum, $r_{\rm kmin}$. (For the single-quark solution above whose $\Phi_k$ minimum is at $r = s/\alpha$ and $\Psi_\lambda^\prime < 0$, the actual minimum $r_m > s/\alpha$).  Succinctly: 
\begin{equation}
\Psi _\lambda ^\prime < 0 \rightarrow d \Phi_k / dr > 0 \rightarrow r _m > r _{\rm kmin}
\end{equation}
\begin{equation}
\Psi _\lambda ^\prime > 0 \rightarrow d \Phi_k / dr < 0 \rightarrow r _m < r _{\rm kmin}
\end{equation}

	To apply the use of (6.20) for determining the minimum energy (for simplicity) we use the single-quark solution (4.11). With $r_m = ns/\alpha $  one obtains the following equation for $n$, ((6.21) applies),
\begin{equation}
2(n+1)^7 (n-1) / n ^3 = + \lambda /k\alpha ^3
\end{equation}
Choosing, for example, the simple values $\lambda/k = 10^{-3}$ say and $\alpha = 10^{-3}, \lambda/k\alpha ^3 = 10^6$, the solution is $n = 12.6$. We would want $r_m$ to be less than one fermi, $10^{-15}m$, so with $r_m =  ns/\alpha < 10^{-15}, s < 10^{-15}\alpha/n$,  thus  
$s < 10^{-19}m$. (For $\lambda /k =1$ and $\alpha = 10^{-2}, n$ is again 12.6 giving $s < 10^{-18}m$).

	A neutral system would require both $P$ and $C$ terms in (6.11). With small enough $K$ there are two possible choices for $P$ and $C$ where the first $\Phi_k > 0$ and the second $\Phi_k < 0$, or vice versa:
\begin{equation}
[P > 0, C < 0: P > -C], \; \mbox{or}\;  [P < 0, C > 0: -P > C]  
\end{equation}
(For positive or negative charged systems appropriate choices for $P$ and $C$ are easily made).		

	 For a 3-quark solution, with the 3rd $\Psi_\lambda ^\prime < 0, \Psi_\lambda $ must have one minimum followed by one maximum.  These two extrema produce two divergences in the central region and thus three extrema in $ \Phi_k$ for three quarks. In region(1) $\Psi_\lambda ^\prime < 0$, in region(2) $\Psi_\lambda ^\prime > 0$ and in region(3) $\Psi_\lambda ^\prime < 0$. For a proton, say, two positive and one negative $\phi_k$ are needed. For a small enough $K$ (6.11) and 
$k > 0$, with $P < 0, C > 0, -P > C$, the first $\phi_k $ is positive, the second is negative and the third is positive, so the energy, $b\phi_k$, of each quark in its region is positive.  Another choice is $P < 0, C< 0$ and $-P > -C$. The two negative slopes represent an up quarks and the one positive slope a down quark, (the $d$ quark always lies be between the $u$ quarks). Formally, $n$ extrema result in 
$n + 1$ quarks with an infinite barrier between adjacent quarks, arranged in `shells' of asymptotically free concentric spherical surfaces, one quark per `shell'. A possible symmetry that groups the quarks into hadrons is discussed in the Symmetry section.

	In the above examples, in each case, the system is completely confined.  With the inclusion of a large enough $K$ term however an unstable quark can result. Choosing $\Psi_\lambda$ with a negative slope for the second quark of a 2-quark particle and choosing $C > 0$ with $P > C$, (the $P$ term changes sign compared to its sign in the first quark) the sum of both $P$ and $C$ terms in (6.11) produce a negative divergence at the $\Psi_\lambda$ maximum and another negative divergence asymptotically. Including now the $K$ term contribution, the second quark's $\Phi_k$ in (6.11) can be written (the $K$ term also changes sign),
\begin{equation}
\Phi_k = - K s^3 \Psi_\lambda /r^2 |\Psi_\lambda ^\prime |- (P-C) /r^2 |\Psi_\lambda^\prime|       
 \end{equation}
With $K < 0$ and $-Ks^3 \Psi_{\lambda \max} > P - C$, the total $\Phi_k$ divergence at $\Psi_{\lambda \max}$ will be positive for a positively charged particle, and asymptotically where the $K$ term $\rightarrow 0$ as $+1/r$, the divergence of $\Phi_k$ will be negative, eliminating absolute confinement. Shown below, solutions exist for the second quark that have a minimum followed by a maximum and then the $-r^2$ divergence, giving only partial confinement. Using (6.25), with a charge $b$ for the second quark, the energy minimum (and maximum) is obtained from the zero derivative of the interaction energy,
\begin{eqnarray}
\xi_k + \xi _\lambda  & = & b ( \phi_k + \psi_\lambda + \, \mbox{const})
\nonumber\\
 &= & bk/s^3 [-Ks^3  \Psi_\lambda  / r^2 |\Psi_\lambda ^\prime| - (P - C)/r^2 |\Psi_\lambda ^\prime |] + b \lambda s^2 \Psi_\lambda  + b \lambda s^2 t 
\end{eqnarray}
where $(P-C) > 0$. (For the first quark the energy will have $+(P + C)$ in (6.26)). To show solutions for an unstable quark we choose $\Psi_\lambda $ from the 2-quark function (6.15). Dividing (6.26) by $bk/s$, putting 
$x = \alpha r/s$, the energy can be expressed in the dimensionless form,
\begin{equation}
E = D (1+ x) /x (3x - 1 ) - U (1+x)^5 /x^2 (3x - 1) + Nx / (1 + x)^4 + T 
\end{equation}
where $D = -\alpha K, U = \alpha^2 (P- C), N = \lambda/\alpha k$ and $T = s^3\lambda t/k$. (At this second quark, the $U$ term confirms that asymptotically, $E$ diverges as $-x^2$). For a positive divergence at $x = 1/3$ (the maximum of $\psi_\lambda$, from (6.16)) $D/U$ must be greater than 9.48. Without loss of generality we put $U = 1$. The zero derivative of (6.27) gives the following relation between $D$ and $N$ for $x > 1/3$,  
\begin{equation}
D = \frac{2(1+x)^4(3x^2-6x+1)}{x(-3x^2 - 6 x +1)} + \frac{(3x - 1)^3 x^2 N}{(1+x)^5(-3x^2 - 6 x + 1)} 
\end{equation}
For a few solutions, Table 1,  we chose, arbitrarily, a specific value for $D$ and a specific value for $x$ to be an extremum; the following values were obtained for $N, x_{\rm min}, x_{\rm max}, E_{\rm min}$ and $E_{\rm max}$.
\begin{table}[h!]
\begin{center}
\caption{Unstable Solutions}
\begin{tabular}{lrrrrrr}
$D$ & $ N$ & $x_{\rm min}$ & $x_{\rm max}$ & $E_{\rm min}$& $E_{\rm max}$& $(E_{\rm max}  - E_{\rm min})$\\
\hline\\
10  &     $-83.4$    &    0.8    &  3/2   &   $-12.3$   &  $ -10.8 $   &     1.5\\
10   &     $-152 $   &     0.7   &   2.0    &   $-17$ 	 &     $ -12.9 $ & 4.1\\   
10   &     $-488$    &     0.5  &    3.0   &    $-49$	  &     $ -18.3 $  &     30.7\\
$10^2  $  &   $-1163$   &    2.0  &    2.5   &   $-10.93$  &  $-10.77$   &   0.16\\
 $10^3 $  &   $-10390$   &  2.3   &   3.0   &    29.1   &    30.7    &    1.6
\end{tabular}
\end{center}
\end{table}
The variation of depths of the various minima, $(E_{\rm max} - E_{\rm min})$, range from about 0.16 to 31, reflecting various lifetimes for the different particles. (The initial chosen extremum turned out to be a maximum except the one for $D = 10^2$). To each set with negative energy at the minimum a value of $T$ must be added to shift to a positive energy. There is no solution with $\gamma = -2$ when $C = P (U = 0)$, for then $\phi_k $ would go as $1/r$ asymptotically.  With the decay of an unstable charged 2-quark particle, 
$\phi_k \rightarrow 0$, leaving $\phi = \phi_p$ and $\psi = \phi_p + \psi_\lambda $. Without confinement the interaction energies $b\phi_k$ and $b \psi_\lambda$ are superseded and the emitted free particle, with unit charge and corresponding equal $\phi_p$ and $\psi_p$ (a possible leptonic decay), would have a mass obtained from (5.2),

\begin{equation}
m = 1/2 \int [ (d \phi _p / d r) ^2 + (d \phi _p / dr)(d \psi _\lambda /dr) ]r^2 dr,
\end{equation}
showing a contribution to the mass from the weak potential $ \psi_\lambda $.

\section{Simple Harmonic Motion}

	It is instructive to examine a quark's simple harmonic motion near the bottom of its energy-well, its asymptotically free spherical surface. The force on a quark, $-d\xi/dr$, for a small 
displacement $\varepsilon$ from $r_m$ at the energy minimum $\xi    (r_m)$ is
\begin{equation}
  F = -\xi ^\prime (r_m+\varepsilon)  =  - \xi ^{\prime \prime}(r_m)\varepsilon   + O(\varepsilon ^2) + \ldots        
\end{equation}	
With mass $m = \xi(r_m)/c^2$, (Coulomb energy is assumed to be small), the equation 
\begin{equation}
md^2 \varepsilon / d t^2 = F 
\end{equation}	
becomes,			
\begin{equation} 
d^2 \varepsilon /dt^2 = - [\xi^{\prime \prime} (r_m ) /\xi (r_m )] c^2 \varepsilon 
\end{equation}
Hence, the angular frequency	
\begin{equation}
\omega = [ \xi^{\prime \prime } (r_m )/\xi(r_m )] ^{1/2} c 
\end{equation}				 

Again, using the single-quark solution (4.11), one finds for the new minimum at $r = ns/\alpha$, 
\begin{equation}
\xi^{\prime \prime }  = 2 b k \alpha ^2 /s^3 [( n + 1 ) ^7 (n^2 - 2 n + 3) \alpha ^2 + 6 n ^4 \lambda /k ] /n ^4 (n + 1) ^5
\end{equation}
\begin{equation}
\xi = bk/s[( n + 1)^7 \alpha ^2 + n ^2 \lambda /k ] /n^2 (n + 1) ^ 3 
\end{equation}
\begin{equation}\xi^{\prime \prime }  / \xi  = \frac{2 \alpha ^2 [( n + 1 )^7 (n ^2  - 2n + 3 ) \alpha ^2 + 6 n ^4 \lambda /k ]}{s^2 [( n + 1 )^7 \alpha ^2 + n^2 \lambda /k ]n^2 (n+1) ^2} 
\end{equation}
Using the values from the example for (6.23) where $\lambda /k = \alpha  = 10^{-3}$ and $n = 12.6$, one finds the frequency from (7.7) and (7.4),
\begin{equation}
\omega \sim 0.1 c \alpha /s = 3 \times 10^4 /s > 3\times 10^{23} / \mbox{sec},
\end{equation}
corresponding to a photonic energy $h\omega/2\pi  \approx 0.2GeV$. 
(For $\lambda/k =1$ and $\alpha = 10^{-2}$ where $n$ is again $12.6, h\omega/2\pi \approx 0.02GeV)$. The $\phi_k$ and $\psi_\lambda$ energies from (7.6) with $n = 12.6$ and $\lambda/k  = 10^{-3}$ plus the Coulomb energy are: 
\begin{equation}
\xi_k \approx 2 \times 10^{-4} bk/s, \;\; \xi_\lambda \approx 4 \times 10^{-6} b \lambda /s = 2 \times 10^{-5} \xi _k, \;\; \xi _{\rm coul} \approx 4 \times 10^{-6} b^2 /s 
\end{equation}
Assuming $k \geq b$, the energy is dominated by $\xi_k$. Writing $b = b^\prime e, k = k^\prime e$ (with $k^\prime $ the amplitude, $ b^\prime$  either 1/3 or 2/3 and $e$ the electronic charge) and $s = 10^{-19}m, \xi_k \sim b^\prime k^\prime 3MeV$. 

\section{Symmetry}

	These quarks consist of products of two equal particular solutions, $\phi_p$ and $\psi_p$, (shown symbolically distinct for clarity---subscript $p$ will be suppressed). With the two different $q$ values $\pm 1/3$ for which the same fractional charges obtain (3.4), (3.5) four different $ \phi \psi $  products can be made, such as $\phi(1/3)\psi(-1/3)$ etc., representing 4 possible quarks (or anti-quarks) with the same charge. For brevity we show an up quark, say, $\{\phi_{2/3}(1/3)\psi_{2/3}(-1/3)\}$ with the $q$-pair $(1/3,  -1/3)$ as $u(+, -)$ similarly for other quarks. The pairing of the $q$ values with the $\phi \psi$ combination implies that there are effectively 4 distinct $q$-pairs $(+,+), (-,-), (+,-), (-,+)$ with the obvious property that the sum of their 8 individual $q$'s is zero; but as we see next, zero sum $q$'s ($q$-neutrality) is a necessary and sufficient condition for the formation of baryons and mesons (evoking QCD's color neutrality).

From the 4 $q$-pairs a baryon's three quarks can always have three $q$-pairs with zero sum $q$'s consisting of the two $q$-pairs $(+,+), (-,-)$ plus, randomly, either the $q$-pair $(+,-)$ or $(-,+)$. A proton, for example, could change continually between $[u(-,-),  u(+,-),  d(+,+)]$ and 
$[u(-,+), u(+,+), d(-,-)]$ (order of $q$-pairs arbitrary). Mesons can also use all 4 $q$-pairs in their sets of [(quark), (anti-quark)] combinations, $[(+,+), (-,-)]$ and $[(+,-), (-,+)]$ plus the two sets with the signs reversed, all sets with zero sum $q$'s. (How $q$-pairs may change is discussed below in Conclusions).
	
	With the bi-functional form of quarks as the basis for the effective  non-commutativity of     $q$-pairs $(+,-)$ and $(-,+)$, the 4 distinct $q$-pairs 
$(+, +) ,(-, -), (+, -), (-, +)$  show that 
$q$-neutrality can occur in 4, 3, 2 and 1 $q$-pairs, with possible relevance to multi-quark particles beyond hadrons as well as single-quark particles; though the latter are consistent with the single-quark solutions obtained above, these have not been observed[1].

\section{Conclusions}

	The Lagrangian (4.2), although missing initially the usual kinetic terms  $F_{\mu \nu}  F_{\mu \nu} $ and $G_{\mu \nu} G_{\mu \nu} $, was restored to the degree that the choice of equal particular solutions of the two field equations merged the corresponding $F_{\mu \nu} $ and $G_{\mu \nu} $. This led to an electromagnetic stress-energy tensor with an energy density corresponding to the Coulomb energy density of a single charge  (5.3)---raising inevitably, the intriguing question of a possible quark structure. 

	The results, such as fractional charges and absolute confinement, including asymptotic freedom, are in accord with those of QCD; as well, the manifest intrinsic symmetry adds support to the model. The $q$-pairs, combined with the bifunctional form of the quarks establish the quarks' symmetry, but are not involved in any direct interaction. 	

	Solutions with multi-quarks are easily obtained describing both stable and unstable particles. The significant result where the strong confining potential $\phi_k$ is completely determined by the short-range weak potential $\psi_\lambda$ highlights the unity between them and electromagnetic potentials, emphasizing also the primacy of the weak potential with respect to the derivative strong confining potential. From the bottom of the energy-well on its spherical surface a quark can move radially in simple harmonic motion. This of course does not preclude a possible additional force-free motion on a spherical surface. 

	How the $q$-pairs in hadrons can continue to change is not, in general, within the purview of these static solutions. However, with expected time dependence, the $q$ values in (1.8) could be viewed as changing rapidly and continually between $1/3$ and $-1/3$, leading to random changes in the $q$-pairs. Classically, this `virtual' process should not be observable; the average $q$ value would be zero and (1.8) would revert to the familiar Maxwell equations. However, because of the success of the static  solutions in describing important quark properties, we suggest that they could be interpreted as displaying `snapshots' of this process. 
	
	Overall, this presents a different and possibly fruitful perspective on quarks within a framework of classical fields where all forces and interactions are electromagnetic.

\section*{References}

\begin{enumerate}
\item  Perl, M.L.  2009.  Nuclear Physics B, {\em Proceedings Supplements}, {\bf 189}, 5.
\end{enumerate}

\section*{Acknowledgement}

	I want to thank J.\ David Jackson for many helpful comments and suggestions, also Werner Israel for useful discussions.

\smallskip

\noindent \rule{4.5in}{.07ex}

\noindent The original publication is available at  
http://www.springerlink.com.

\end{document}